\documentclass[prd,twocolumn,amsmath,amssymb,nofootinbib,floatfix,superscriptaddress]{revtex4}

\usepackage{graphicx,bm,float}

\makeatletter
\def\graphicscale{\twocolumn@sw{0.3}{0.4}}
\def\graphicthreescale{\twocolumn@sw{0.3}{0.4}}

\begin{document}

\title{Berezinskii-Kosterlitz-Thouless transitions in \\
two-dimensional lattice SO($N_c$) gauge theories \\ 
with two scalar flavors}

\author{Claudio Bonati} 
\affiliation{Dipartimento di Fisica dell'Universit\`a di Pisa 
       and INFN, Pisa, Italy}

\author{Alessio Franchi} 
\affiliation{Dipartimento di Fisica dell'Universit\`a di Pisa,
       Pisa, Italy}

\author{Andrea Pelissetto}
\affiliation{Dipartimento di Fisica dell'Universit\`a di Roma Sapienza
        and INFN, Roma, Italy}

\author{Ettore Vicari} 
\affiliation{Dipartimento di Fisica dell'Universit\`a di Pisa
       and INFN, Pisa, Italy}

\date{\today}

\begin{abstract}
We study the phase diagram and critical behavior of a two-dimensional
lattice SO($N_c$) gauge theory ($N_c \ge 3$) with two scalar flavors,
obtained by partially gauging a maximally O($2N_c$) symmetric scalar
model. The model is invariant under local SO($N_c$) and global O(2)
transformations.  We show that, for any $N_c \ge 3$, it undergoes
finite-temperature Berezinskii-Kosterlitz-Thouless (BKT) transitions,
associated with the global Abelian O(2) symmetry. The transition
separates a high-temperature disordered phase from a low-temperature
spin-wave phase where correlations decay algebraically (quasi-long
range order).  The critical properties at the finite-temperature BKT
transition and in the low-temperature spin-wave phase are determined
by means of a finite-size scaling analysis of Monte Carlo data.

\end{abstract}

\maketitle


\section{Introduction}
\label{intro}

Abelian and non-Abelian gauge symmetries appear in various physical
contexts. For instance, they are relevant for the theories of
fundamental interactions~\cite{Wilson-74,ZJ-book,Weinberg-book} and in
the description of some emerging phenomena in condensed matter
physics~\cite{ZJ-book,Sachdev-19,Anderson-book}.  The main features of
these theories, such as the spectrum, the phase diagram, and the
critical behavior at thermal and quantum transitions, crucially depend
on the interplay between global and local gauge symmetries.

These issues have been recently investigated in several
two-dimensional (2D) lattice gauge models, considering: (i) the
multicomponent lattice Abelian-Higgs model~\cite{BPV-19-ah2},
characterized by a global SU($N_f$) symmetry ($N_f\ge 2$) and a local
U(1) gauge symmetry; (ii) the multiflavor lattice scalar quantum
chromodynamics~\cite{BPV-20-qcd2}, characterized by a global SU($N_f$)
symmetry and a local SU($N_c$) gauge symmetry; (iii) lattice SO($N_c$)
gauge models with $N_f\ge 3$ real scalar flavors~\cite{BFPV-20-ong},
characterized by a non-Abelian O($N_f$) global symmetry.  In agreement
with the Mermin-Wagner theorem \cite{MW-66}, all these 2D lattice
gauge models do not have finite-temperature transitions.  A critical
behavior is only observed in the zero-temperature limit: for $T\to 0$
the correlation length increases exponentially, as in the 2D O($N$)
$\sigma$ model wih $N\ge 3$ and in the 2D CP$^{N-1}$ model with $N\ge
2$~\cite{ZJ-book}.  The interplay of global non-Abelian symmetries and
local gauge symmetries determines the large-scale properties of the
system in the zero-temperature limit, and therefore, the field theory
realized in the corresponding continuum limit.

The results for the above-mentioned lattice gauge models support the
following general conjecture, originally put forward in
Ref.~\cite{BPV-20-qcd2}.  The universal features, i.e., the
universality class, of the asymptotic low-temperature behavior of
lattice gauge models is the same as that of the 2D field theories
defined on the symmetric spaces~\cite{BHZ-80,ZJ-book} that have the
same global symmetry.  According to this conjecture, the
zero-temperature critical behavior of multiflavor Abelian-Higgs models
and lattice scalar chromodynamics with $N_f$ scalar flavors belongs to
the universality class of the 2D CP$^{N_f-1}$ model, as both models have
the same global SU($N_f$) symmetry.  Analogously, lattice SO($N_c$)
gauge theories with $N_f\ge 3$ real scalar flavors have the same
critical behavior as RP$^{N_f-1}$ models~\cite{BFPV-20} with the same
global O($N_f$) symmetry.  These predictions have been numerically
verified in Refs.~\cite{BPV-19-ah2,BPV-20-qcd2,BFPV-20-ong}.  We note
that all cases considered so far involve systems with global
non-Abelian symmetries, which are not expected to show
finite-temperature transitions in two dimensions \cite{MW-66}.

In this paper we investigate 2D lattice non-Abelian gauge models that
undergo a finite-temperature transition, and show that also in this
case the conjecture holds.  We consider a 2D lattice SO($N_c$) gauge
model with two real scalar flavors, obtained by partially gauging a
maximally O($2N_c$) symmetric scalar theory. For $N_c\ge 3$ this model
is characterized by a global Abelian O(2) symmetry (for $N_c=2$ the
global symmetry group turns out to be SU(2) \cite{BFPV-20-ong}, which
is non-Abelian, and therefore we only expect a zero-temperature
critical behavior).  If the general conjecture extends to systems with
global Abelian symmetries, we expect this model to have the same
critical behavior as the O(2)-invariant $XY$ lattice model. Therefore,
for $N_c\ge 3$, 2D lattice SO($N_c)$ gauge models with two scalar
flavors may undergo a finite-temperature
Berezinskii-Kosterlitz-Thouless (BKT)
transition~\cite{KT-73,Berezinskii-70,Kosterlitz-74,JKKN-77,
  HMP-94,HP-97,Balog-01,Hasenbusch-05,PV-13}, between the
high-temperature disordered phase and a low-temperature spin-wave
phase characterized by quasi-long range order (QLRO) with vanishing
magnetization. We recall that BKT transitions are characterized by an
exponentially divergent correlation length $\xi$ at a finite critical
temperature.  Indeed, $\xi$ behaves as $\xi\sim \exp(c/\sqrt{T-T_c})$
approaching the BKT critical temperature $T_c$ from the
high-temperature phase.

To verify the general conjecture for the lattice SO($N_c$) gauge
theory with two scalar flavors, we present finite-size scaling (FSS)
analyses of Monte Carlo simulations for several $N_c\ge 3$. We
anticipate that the numerical results confirm the presence of a
low-temperature QLRO phase, separated by a BKT transition from the
high-temperature disordered phase. These results extend the validity
of the conjecture to 2D lattice non-Abelian gauge theories with global
Abelian symmetries.

The paper is organized as follows. In Sec.~\ref{model} we define the
lattice SO($N_c$) gauge model with scalar fields, and the
gauge-invariant observables that we consider in our numerical
study. We also describe the FSS analysis we use to investigate the
phase diagram and to determine the nature of the critical behavior.
Sec.~\ref{numres} reports the numerical results for $N_c=3,4,5$. We
show that QLRO holds in the low-temperature phase and that the
transition between the high-temperature and the low-temperature QLRO
phase is a BKT one, as in the standard XY model.  Finally, in
Sec.~\ref{conclu} we summarize and draw our conclusions.

\section{2D lattice SO($N_c$) gauge models}
\label{model}

We consider a multiflavor lattice SO($N_c$) gauge model defined on
square lattices of linear size $L$ with periodic boundary conditions.
It is obtained~\cite{BPV-20-on} by partially gauging a maximally
O($M$) symmetric model with $M=N_f N_c$, defined in terms of real
unit-length matrix variables $\phi_{\bm x}^{af}$, with $a=1,..,N_c$
and $f=1,...,N_f$ (we will refer to these two indices as {\em color}
and {\em flavor} indices, respectively), such that ${\rm
  Tr}\,\phi_{\bm x}^t \phi_{\bm x} = 1$.  Using the Wilson
approach~\cite{Wilson-74}, we introduce gauge variables associated
with each link of the lattice. The Hamiltonian reads~\cite{BPV-20-on}
\begin{eqnarray}
H = -  N_f \sum_{{\bm x},\mu} {\rm Tr} \,\phi_{\bm x}^t \, V_{{\bm
    x},\mu} \, \phi_{{\bm x}+\hat{\mu}} - {\gamma\over N_c} \sum_{{\bm
    x}} {\rm Tr}\, \Pi_{\bm x}\,,\;
\label{hgauge}
\end{eqnarray}
where $V_{{\bm x},\mu} \in {\rm SO(}N_c)$, $\Pi_{\bm x}$ is the
plaquette operator
\begin{equation}
\Pi_{\bm x}= V_{{\bm x},1} \,V_{{\bm x}+\hat{1},2} \,V_{{\bm
    x}+\hat{2},1}^t \,V_{{\bm x},2}^t \,.
\label{plaquette}
\end{equation}
We set the lattice spacing equal to 1, so that all lengths are
measured in units of the lattice spacing.  The plaquette parameter
$\gamma$ plays the role of inverse gauge coupling. The partition
function reads
\begin{eqnarray}
Z = \sum_{\{\phi,V\}} e^{-\beta \,H}\,,\qquad \beta\equiv 1/T\,.
\label{partfun}
\end{eqnarray}
Note that, for $\gamma\to\infty$, the link variables $V_{\bm x}$
become equal to the identity modulo gauge transformations. Thus, one
recovers the O($M$)-symmetric nearest-neighbor $M$-vector model, which
does not have a finite-temperature transition and becomes critical
only in the zero-temperature limit~\cite{ZJ-book,PV-02}.

For $N_c\ge 3$ the global symmetry group of model (\ref{hgauge}) is
O($N_f$). For $N_c=2$ the global symmetry is actually
larger~\cite{BPV-20-on}, since the model can be exactly mapped onto
the two-component lattice Abelian-Higgs model, which is invariant
under local U(1) and global SU$(N_f)$ transformations. Therefore, for
$N_f=N_c=2$ the model has a zero-temperature critical behavior
belonging to the universality class of the CP$^1$ field
theory~\cite{BPV-19-ah2}, which is equivalent to that of the nonlinear
O(3) $\sigma$ model.  In the following we consider only the case $N_c
\ge 3$.

For $N_f=2$ and $N_c\ge 3$ the theory is characterized by a global
Abelian O(2) symmetry. The conjecture we have discussed in the
introduction suggests therefore that the two-flavor gauge model has a
finite-temperature transition analogous to that occurring in 2D
O(2)-invariant spin models, which undergo a BKT transition from the
disordered phase to a low-temperature QLRO phase~\cite{ZJ-book}.  As
we shall see, this conjecture is supported by the numerical results.

To determine the nature of the transitions, we will perform a FSS
analysis~\cite{FB-72,Barber-83,Privman-90,PV-02} of the numerical
data.  We focus on the correlations of the gauge-invariant bilinear
operator
\begin{equation}
Q_{\bm x}^{fg} = \sum_a \phi_{\bm x}^{af} \phi_{\bm x}^{ag} - 
    {1\over 2} \delta^{fg}\,.
\label{qdef}
\end{equation}
Note that, for $N_f=2$, $Q_{\bm x}$ has only two independent real
components.  We consider the two-point function
\begin{equation}
G({\bm x}-{\bm y}) = \langle {\rm Tr}\, Q_{\bm x} Q_{\bm y} \rangle\,,  
\label{gxyp}
\end{equation}
where the translation invariance of the system has been taken into
account. We define the susceptibility $\chi=\sum_{\bm x} G({\bm x})$
and the correlation length
\begin{eqnarray}
\xi^2 = {1\over 4 \sin^2 (\pi/L)} {\widetilde{G}({\bm 0}) -
  \widetilde{G}({\bm p}_m)\over \widetilde{G}({\bm p}_m)}\,,
\label{xidefpb}
\end{eqnarray}
where $\widetilde{G}({\bm p})=\sum_{{\bm x}} e^{i{\bm p}\cdot {\bm x}}
G({\bm x})$ is the Fourier transform of $G({\bm x})$, and ${\bm p}_m =
(2\pi/L,0)$.  We also consider universal RG invariant quantities, such
as the Binder parameter $U$
\begin{equation}
U = {\langle \mu_2^2\rangle \over \langle \mu_2 \rangle^2} \,, \qquad
\mu_2 = {1\over V^2}
\sum_{{\bm x},{\bm y}} {\rm Tr}\,Q_{\bm x} Q_{\bm y}\,,
\label{binderdef}
\end{equation}
where $V=L^2$ (note that $\chi=V\langle \mu_2\rangle$), and the ratio
\begin{equation}
R_\xi\equiv \xi/L\,.
\label{rxidef}
\end{equation}
In the FSS limit we have (see, e.g., Ref.~\cite{BPV-19-ah2})
\begin{equation}
U(\beta,L) \approx F_U(R_\xi)\,,
\label{r12sca}
\end{equation}
where $F_U(x)$ is a universal scaling function that completely
characterizes the universality class of the transition.  In
particular, universality is expected at BKT transitions and in the
whole low-temperature spin-wave phase, see, e.g.,
Refs.~\cite{HMP-94,HP-97,Hasenbusch-08,PV-13,CNPV-13,DV-17}.

Because of the universality of relation (\ref{r12sca}), we use the
plots of $U$ versus $R_\xi$ to identify the models that have the same
universal behavior.  If the estimates of $U$ for two different systems
fall onto the same curve when plotted versus $R_\xi$, the transitions
in the two models belong to the same universality class.  Therefore,
we will compare the FSS curves for the lattice SO($N_c$) gauge model
with the analogous ones for the 2D $XY$ model. If the data for the two
models have the same scaling behavior, we will conclude that the gauge
model undergoes a BKT transition as the XY model.  The same strategy
was employed in Refs.~\cite{BPV-19-ah2,BPV-20-qcd2,BFPV-20-ong}, to
characterize the asymptotic zero-temperature behavior of 2D lattice
gauge models with non-Abelian global symmetry group.

\section{Numerical results}
\label{numres}

\subsection{The conjecture for systems with O(2) global symmetry}
\label{wohy}

We wish to verify numerically the general conjecture originally put
forward in Ref.~\cite{BPV-20-qcd2}. In the present case it predicts
that, for any $N_c \ge 3$, the lattice model with Hamiltonian
(\ref{hgauge}) with two flavors undergoes a transition analogous to
that of the paradigmatic 2D O(2) invariant $XY$ model defined by the
Hamiltonian
\begin{equation}
H_{XY} = -  \sum_{{\bm x},\mu}  {\rm Re}\,\psi_{\bm x}^* \, 
\psi_{{\bm x}+\hat{\mu}} \, ,
\label{xymodel}
\end{equation}
where $\psi_{\bm x}$ are complex phase variables, $|\psi_{\bm x}| =
1$, associated with each site of the square lattice.  This model
undergoes a BKT transition at
$\beta_c=1.1199(1)$~\cite{HMP-94,Hasenbusch-05}, with a
low-temperature phase that shows QLRO with vanishing magnetization.

The correspondence can be justified using the arguments presented in
Ref.~\cite{BFPV-20-ong}. If the conjecture holds, the lattice model
(\ref{hgauge}) with $N_f$ scalar flavors should be related to the 2D
RP$^{N_f-1}$ model, defined by the Hamiltonian
\begin{eqnarray}
H_{\rm RP} = - t \sum_{{\bm x},\mu} (\varphi_{\bm x} \cdot
\varphi_{{\bm x}+\hat\mu})^2 \,,
\label{rpnmodel}
\end{eqnarray} 
where $\varphi_{\bm x}$ is a unit-length $N_f$-component real field.
Indeed, the RP$^{N-1}$ space is a symmetric space that has the same
global O($N_f$) symmetry. The model has also a local ${\mathbb Z}_2$
symmetry, which effectively appears because the order parameter
$Q_{\bm x}$ is invariant under the local ${\mathbb Z}_2$
transformations $\phi_{\bm x} \to s_x \phi_{\bm x}$, $s_x = \pm 1$.
In the RP$^{N-1}$ model the order parameter is
\begin{equation}
q_{\bm
  x}^{fg} = \varphi_{\bm x}^f \varphi_{\bm x}^g - {1\over N_f} \delta^{fg}\,,
 \label{qdefrp}
 \end{equation} 
which is the counterpart of $Q_{\bm x}^{fg}$ defined in the lattice
SO($N_c$) gauge theory.  In the two-flavor case, $N_f=2$, one can
easily show that, for the computation of ${\mathbb Z}_2$
gauge-invariant quantities, the RP$^1$ model can be mapped onto the
$XY$ model. Under this mapping, the order parameter $q_{\bm x}^{fg}$
(which has only two independent real components) is mapped onto the
complex field $\psi_{\bm x}$ of the $XY$ model. Therefore, the
critical behavior of the correlation function of the operator $Q_{\bm
  x}$, defined in Eq.~(\ref{gxyp}), is expected to correspond to that
of the two-point function
\begin{equation}
G_{XY}({\bm x},{\bm y}) = \langle \psi_{\bm x}^* \psi_{\bm
  y}\rangle\,,
\label{gxy}
\end{equation}
in the $XY$ model.  Using $G_{XY}$, one can then define the
correlation length $\xi$, the Binder parameter $U$, and the ratio
$R_\xi$, using again Eqs.~(\ref{xidefpb}), (\ref{binderdef}), and
(\ref{rxidef}), respectively.

\subsection{Monte Carlo simulations}
\label{MCsim}

In the following we report numerical results for the 2D lattice
SO($N_c$) gauge theories with two scalar flavors,
cf. Eq.~(\ref{hgauge}).  We consider square lattices of linear size
$L$ with periodic boundary conditions.  To update the gauge fields we
use an overrelaxation algorithm implemented \emph{\`a la}
Cabibbo-Marinari \cite{Cabibbo:1982zn}, considering the SO(2)
subgroups of SO($N_c$). We use a combination of Metropolis updates and
microcanonical steps \cite{Creutz:1987xi} in the ratio 3:7.  In the
Metropolis update, link variables are randomly generated, and then
accepted or rejected by a Metropolis step \cite{Metropolis:1953am},
with an acceptance rate of approximately 30\%.  For the scalar fields
a combination of Metropolis and microcanonical updates is used, with
the Metropolis step tuned to have an acceptance rate of approximately
30\%.  Errors are estimated using a standard blocking and jackknife
procedure, to take into account autocorrelations, which are expected
to increase roughly as $L^2$.  Typical statistics of our runs, for a
given value of the parameters and of the size of the lattice, are
approximately 10$^7$ lattice sweeps (in a sweep we update once all
lattice variables). For the larger lattice sizes the autocorrelation
times of the observables considered were of order 10$^4$ sweeps at
most, even at $T_c$, thus obtaining a sufficiently large number of
independent measures.

\subsection{The low-temperature spin-wave phase}
\label{swphase}

\begin{figure}[tbp]  
  \includegraphics[width=0.95\columnwidth, clip]{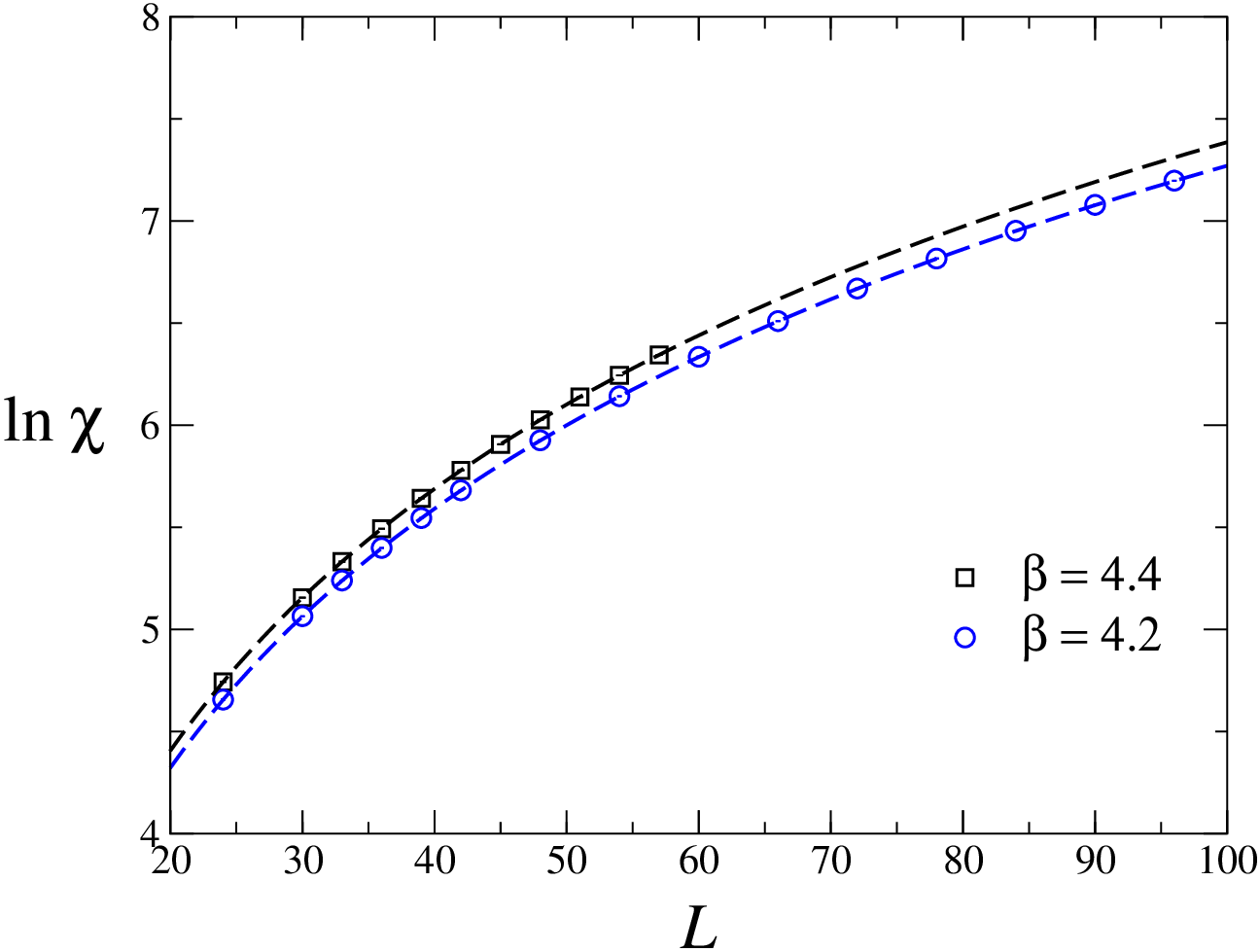}
  \includegraphics[width=0.95\columnwidth, clip]{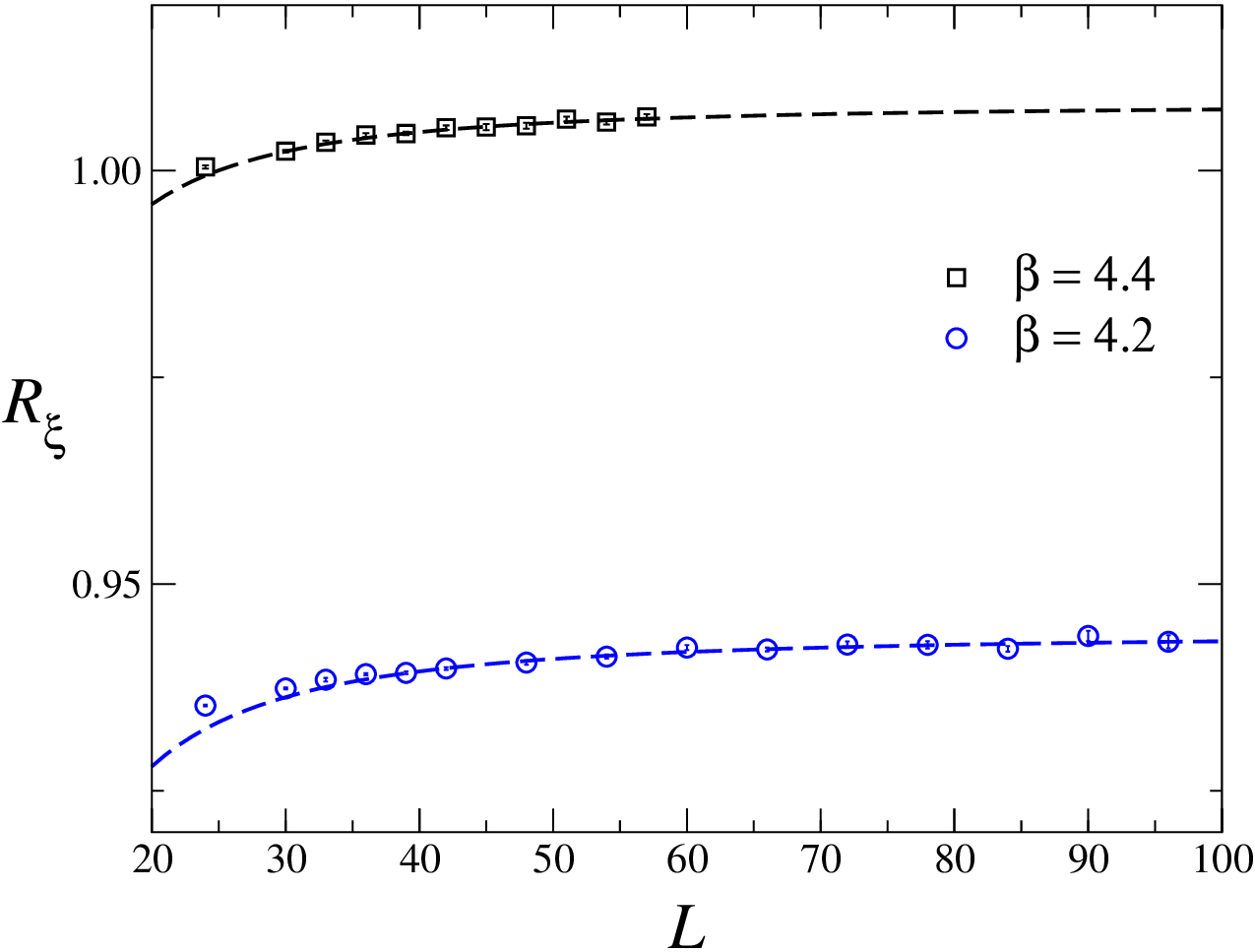}
  \caption{Data of $R_\xi$ (bottom) and $\ln\chi$ (top)
    in the low-temperature spin-wave phase of the
    model with $N_c=3$ and $\gamma=0$, at $\beta=4.2$ and
    $\beta=4.4$. The dashed lines are obtained by fitting the data 
    (results for the smallest lattice sizes have been discarded)
    to the Ans\"atze (\ref{fitchilt}) and (\ref{fitXlt}).
    }
\label{qlrofit}
\end{figure}

\begin{figure}[tbp]  
\includegraphics[width=0.95\columnwidth, clip]{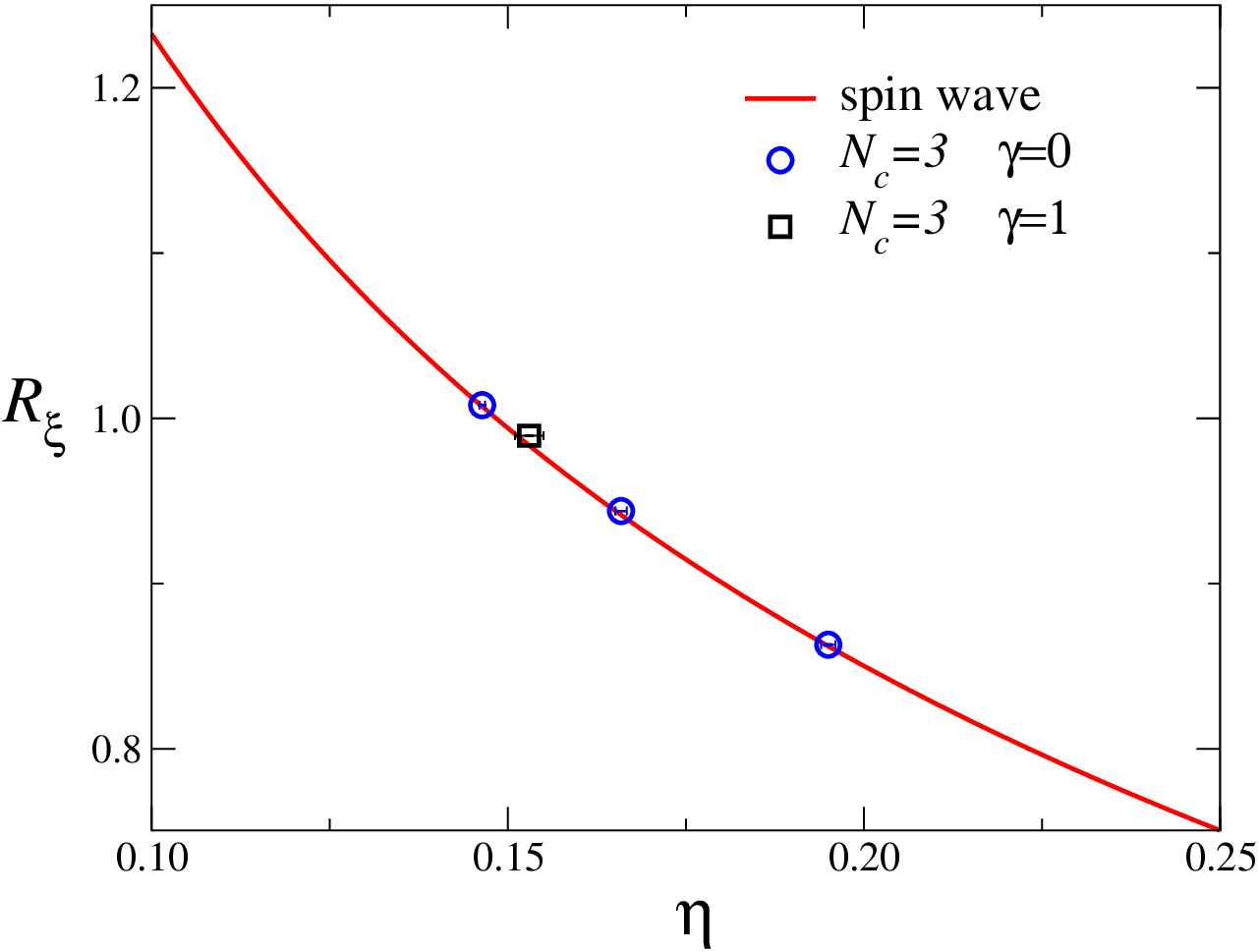}
\caption{Plot of the large-$L$ extrapolations of $R_\xi$ versus $\eta$
  (computed from the finite-size behavior of the susceptibility
  $\chi$) for the lattice SO(3) gauge model. We report results for
  $\gamma=0$ and $\beta=4.0,\,4.2,\,4.4$, and for $\gamma=1$,
  $\beta=4.4$. We also report the universal asymptotic large-$L$ curve
  (full line) computed in the spin-wave theory, for a system with
  square geometry and periodic boundary
  conditions~\cite{CNPV-13,Hasenbusch-05}.  }
\label{qlro}
\end{figure}

To gain evidence of the existence of a low-temperature QLRO phase, we
show that spin-wave relations hold asymptotically for sufficiently low
temperatures.  The spin-wave theory is expected to describe the
critical behavior of the $XY$ model along the line of fixed points
that runs from $T=0$ up to the BKT point $T_c$.  Conformal field
theory, see, e.g., Ref.~\cite{CFT-book}, exactly provides the
large-$L$ limit of the two-point function in the spin-wave model. In
particular, it allows us to compute the universal asymptotic relation
between the ratio $R_\xi$ and the exponent $\eta$. Results for square
lattices with periodic boundary conditions are reported in
Refs.~\cite{CNPV-13,Hasenbusch-05} (see, in particular, the formulas
reported in App.~B of Ref.~\cite{CNPV-13}).  The exponent $\eta$
characterizes the temperature-dependent power-law decay of the
two-point function in the QLRO phase
\begin{equation}
G({\bm x})\sim |{\bm x}|^{-\eta(T)}\,.
\label{gsw}
\end{equation}
Alternatively, we can define it by considering the large-$L$ behavior
of the susceptibility
\begin{equation}
\chi(L,T) \sim L^{2-\eta(T)}.
\end{equation}
In the QLRO phase, $\eta(T)$ varies from $\eta(T_c)=1/4$ to
$\eta(T\to 0) \to 0$, and $R_\xi$ from $R_\xi(T_c) = 0.750691...$ to
$R_\xi(T\to 0)\to\infty$.

We recall that, at $T_c$, the RG theory appropriate for the BKT
transition predicts the asymptotic large-$L$
behavior~\cite{Hasenbusch-05,PV-13,CNPV-13}
\begin{eqnarray}
&&R_\xi(L,T_c) = R_\xi(T_c) + {C_{R_\xi}\over w(L)}
  + O(w^{-2})\,,
\label{rasym}\\
&&w(L) = \ln {L\over \Lambda} + {1\over2} \ln\ln {L\over \Lambda}\,,
\nonumber
\end{eqnarray}
where $\Lambda$ is a model-dependent constant, and $R_\xi(T_c)$ and
$C_{R_\xi}$ are universal. Using the spin-wave theory, one obtains
$R_\xi(T_c) = 0.750691...$ and $C_{R_\xi} = 0.212431...$.  Analogous
results can be obtained for the Binder parameter
$U$~\cite{Hasenbusch-08}, in particular $U(T_c)=1.018192(6)$.

To study the low-temperature behavior, we have performed simulations
for $N_c=3$ at values of $\beta$ such that $R_\xi > R_\xi(T_c)$, using
periodic boundary conditions. We have determined the large-$L$
extrapolations of $R_\xi$ and $\eta$, by fitting the data of $\chi$
and $R_\xi$ at fixed $\beta$ to the Ans\"atze
\begin{eqnarray}
&&\ln\chi(L) = a + (2-\eta)\ln L + b
  L^{-\varepsilon}, \label{fitchilt}\\ 
&& R_\xi(L) = R_\xi + a
  L^{-\varepsilon}, \label{fitXlt}
\end{eqnarray}
respectively, where $\varepsilon$ is the exponent associated with the
expected leading corrections~\cite{CNPV-13,HPV-05}: 
\begin{equation}
\varepsilon={\rm
  Min}(2-\eta,\omega)\,,\quad \omega=1/\eta-4 + O[(1/\eta-4)^2]\,.
  \label{varep}
\end{equation}

For $N_c=3$ and $\gamma=0$ the quality of the fits can be assessed
from the results shown in Fig.~\ref{qlrofit}.  Fits to
Eqs.~(\ref{fitchilt}) and (\ref{fitXlt}) are very good, as it also
supported by the values of $\chi^2/{\rm dof}$ ($\chi^2$ is here the
sum of the fit residuals and dof is the number of degrees of freedom
of the fit), which are smaller than 1, if a few results for the
smallest lattice sizes are discarded.  For $N_c=3$ and $\gamma=0$ we
obtain the large-$L$ extrapolations $\eta=0.195(1),\,0.1659(8),\,
0.1464(4)$, and $R_\xi=0.8630(5),\,0.9439(2),\,1.0080(2)$, for
$\beta=4.0,\,4.2,\,4.4$, respectively.  We have also performed a
detailed study for $\gamma=1$ and $\beta=4.4$.  We obtain
$\eta=0.153(2)$ and $R_\xi=0.9895(3)$.  Note that $\omega$, see
Eq.~(\ref{varep}), is known precisely only for $\eta$ close to 1/4. In
the fits we use $\omega$ as obtained from Eq.~(\ref{varep}), and
therefore $\omega$ gives the leading correction-to-scaling exponent
for $\eta \gtrsim 0.17$.  In such cases, to estimate the error due to
the uncertainty on $\omega$, we checked the variation of the results
of the fits when varying $\omega$ in around the approximation obtained
from Eq.~(\ref{varep}), within a reasonable interval of about 10\%.
This has allowed us to estimate how $\eta$ and $R_\xi$ vary with
changes of $\omega$. Such variation has been included in the final
error.

In Fig.~\ref{qlro} we plot R$_\xi$ versus $\eta$ together with the
universal curve computed in the spin-wave theory.  The results for
$R_\xi$ and $\eta$ are in excellent agreement with the spin-wave
predictions.  This shows the existence of a low-temperature phase with
QLRO, analogous to that occurring in the $XY$ model.

\subsection{FSS at the  BKT transition}
\label{bktres}

\begin{figure}[tbp]  
\includegraphics[width=0.95\columnwidth, clip]{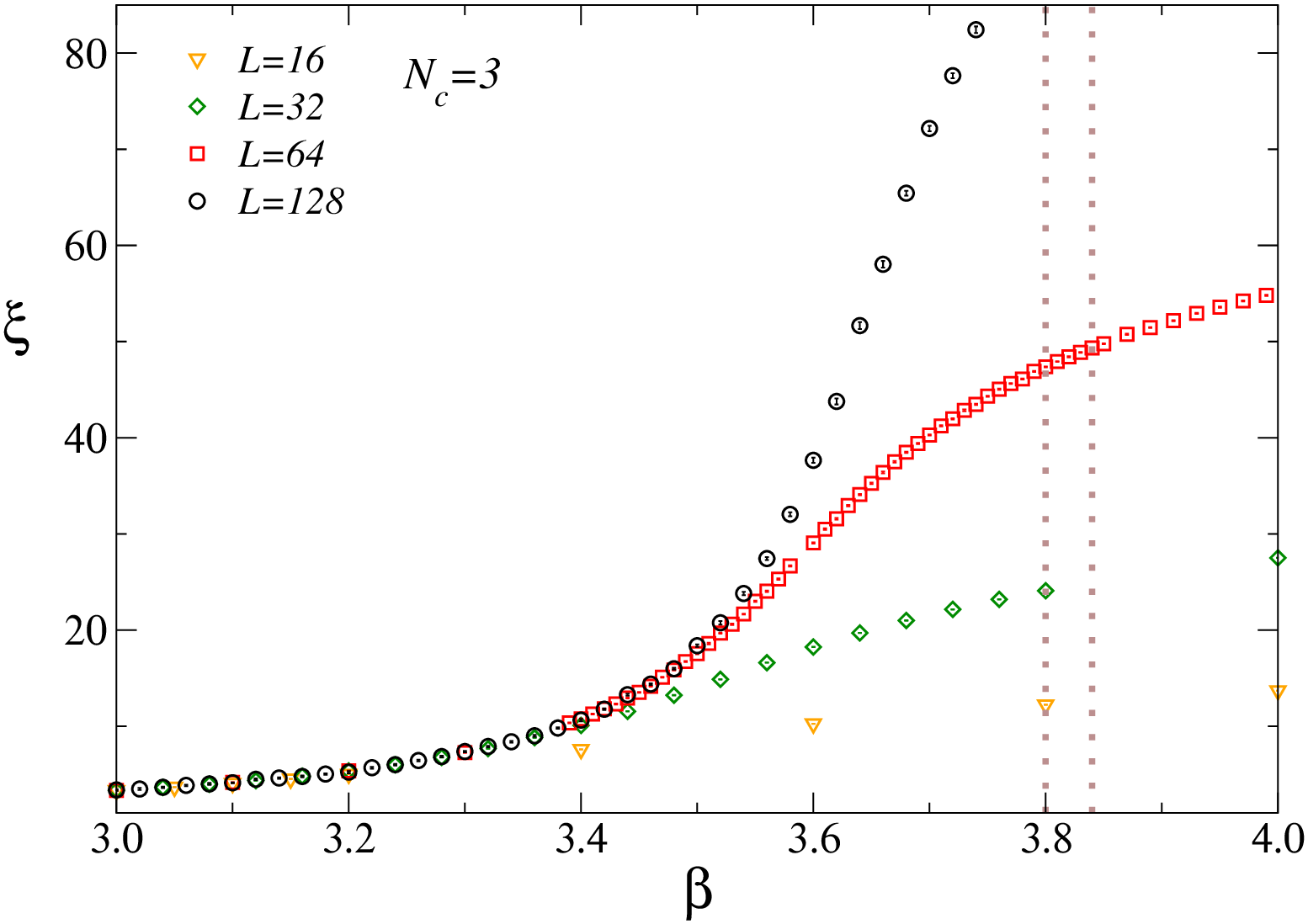}
\caption{Estimates of the correlation length $\xi$ versus $\beta$ for
  the lattice SO(3) gauge model (\ref{hgauge}) with $\gamma=0$, for
  several values of $L$, up to $L=128$.  When the results for
  different values of $L$ agree, they can be considered as good
  approximations of the infinite-volume correlation length, within
  errors.  The vertical lines indicate the interval of values of
  $\beta$ in which the BKT transition occurs.  }
\label{xibetanc3}
\end{figure}

In Sec.~\ref{swphase} we showed that the SO(3) gauge model has a
low-temperature phase with the same features of the low-temperature
phase of the $XY$ model. Now, we focus on the finite-temperature
transition that ends the high-temperature phase, to check whether the
FSS behavior is the same as that observed at the BKT transition of the
$XY$ model.

\begin{figure}[tbp]  
\includegraphics[width=0.95\columnwidth, clip]{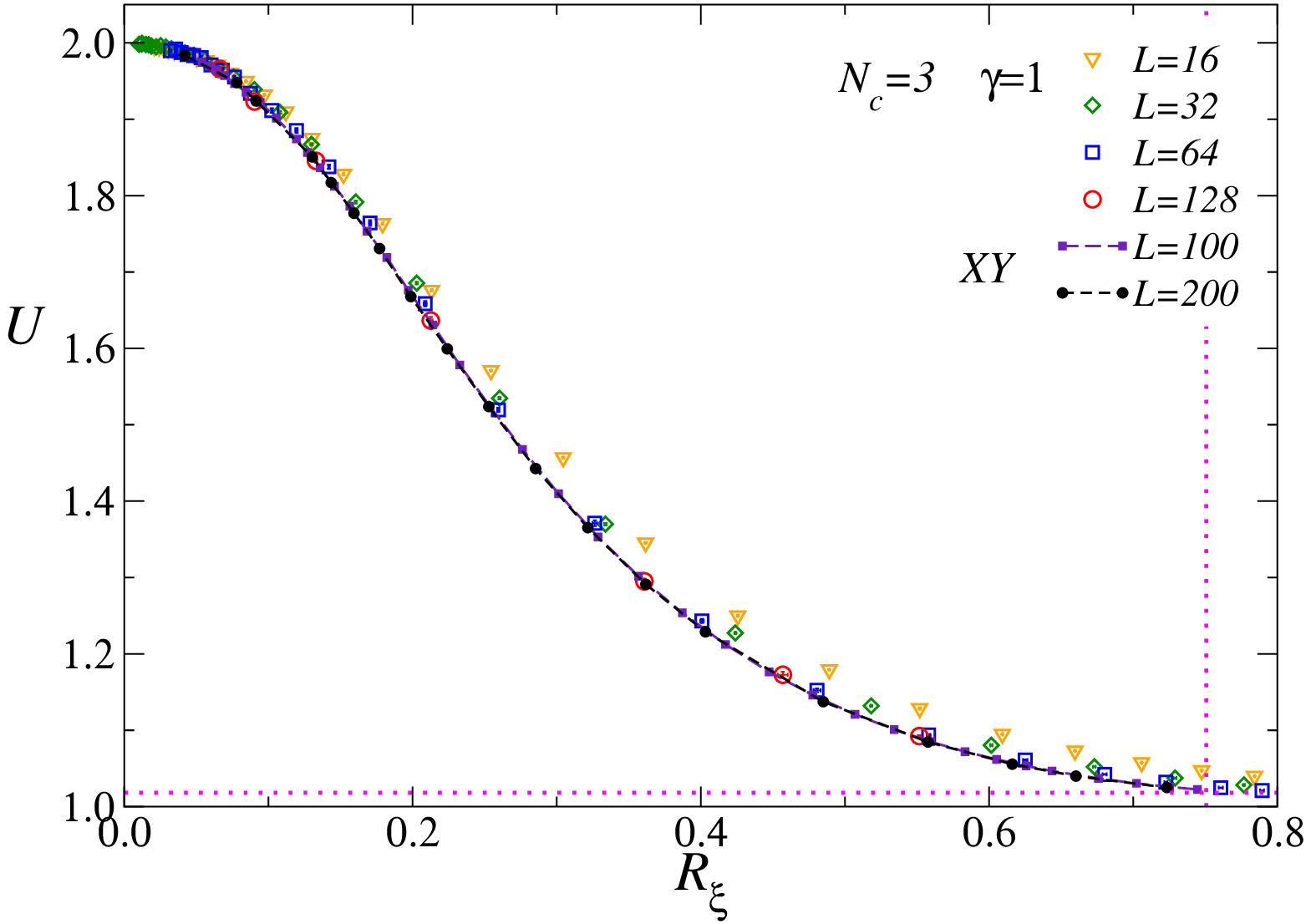}
\includegraphics[width=0.95\columnwidth, clip]{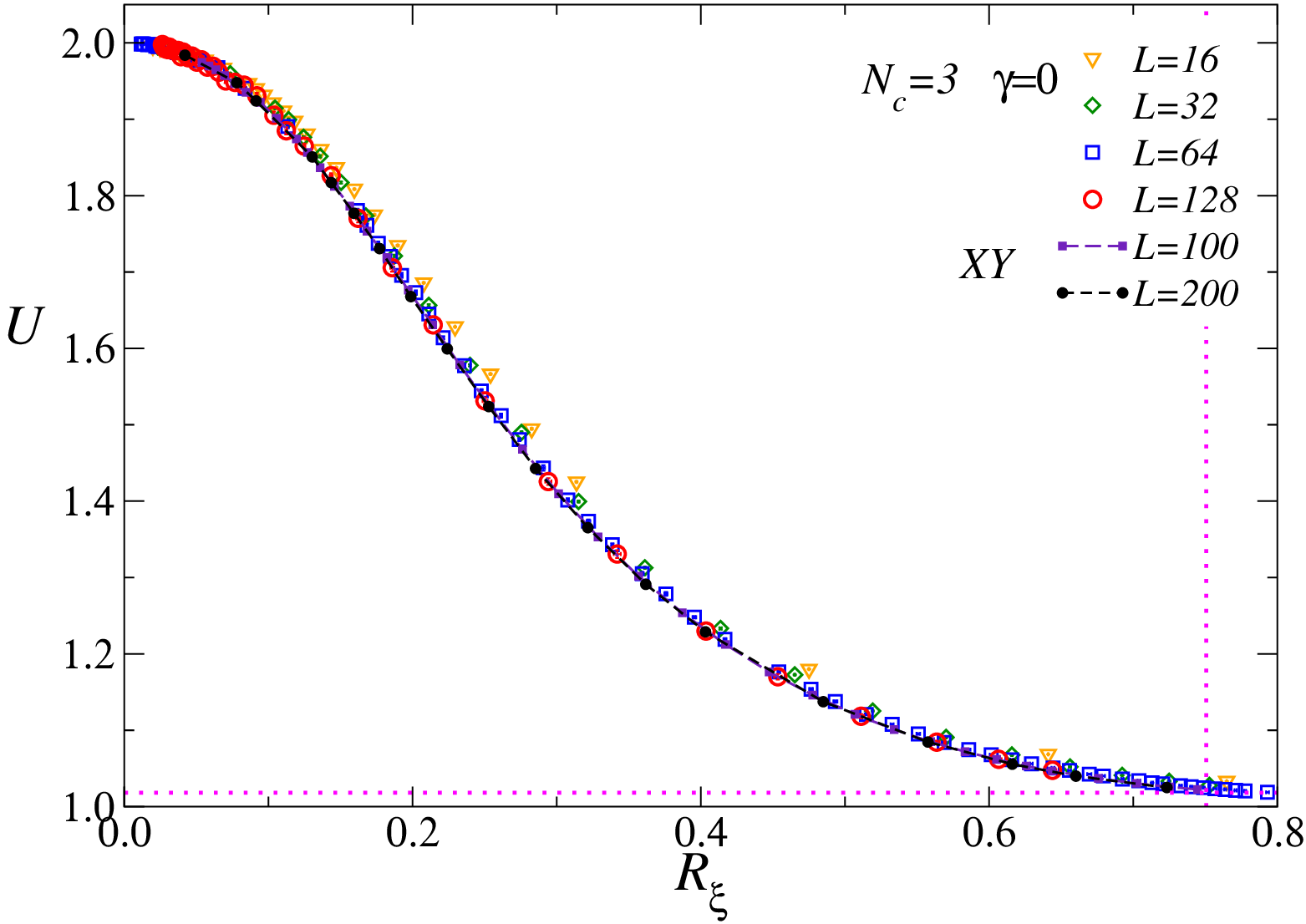}
\includegraphics[width=0.95\columnwidth, clip]{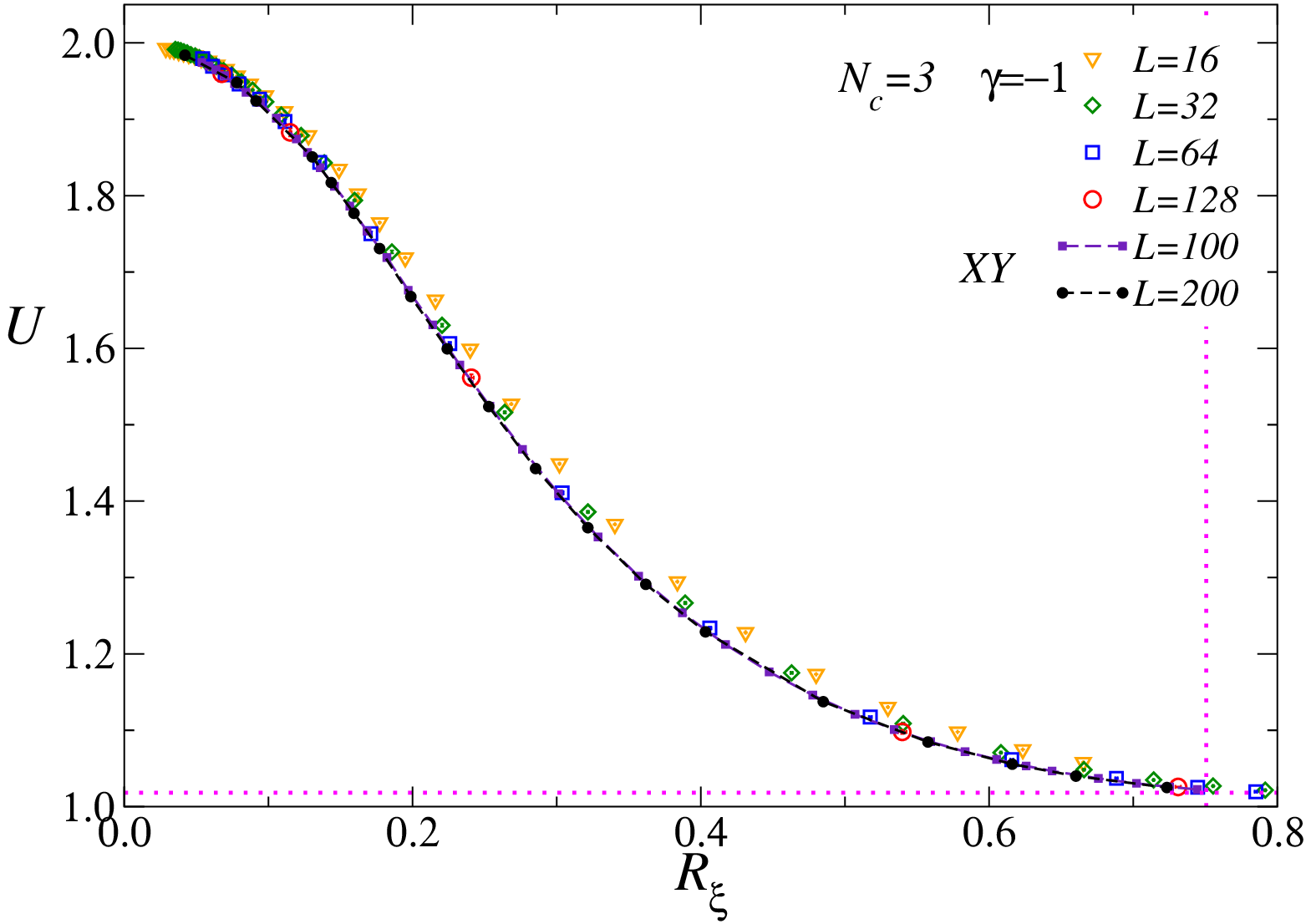}
  \caption{We plot data of $U$ versus $R_\xi$ for $N_c=3$, $\gamma=1$
    (top), $\gamma=0$ (middle), and $\gamma=-1$ (bottom).  We report
    analogous data for the 2D $XY$ model (\ref{xymodel}).  We observe
    a nice agreement, supporting the conjecture that the lattice
    SO($N_c$) gauge model with two scalar flavors undergoes a
    finite-temperature BKT transition for generic values of $\gamma$.
    The horizontal and vertical lines indicate the universal values of
    $U$ and $R_\xi$ at the BKT transition, i.e.  $U(T_c)=1.018192(6)$
    and $R_\xi(T_c) = 0.750691...$,
    respectively~\cite{Hasenbusch-05,Hasenbusch-08}.  }
  \label{urxi3}
\end{figure}

\begin{figure}[tbp]  
\includegraphics[width=0.95\columnwidth, clip]{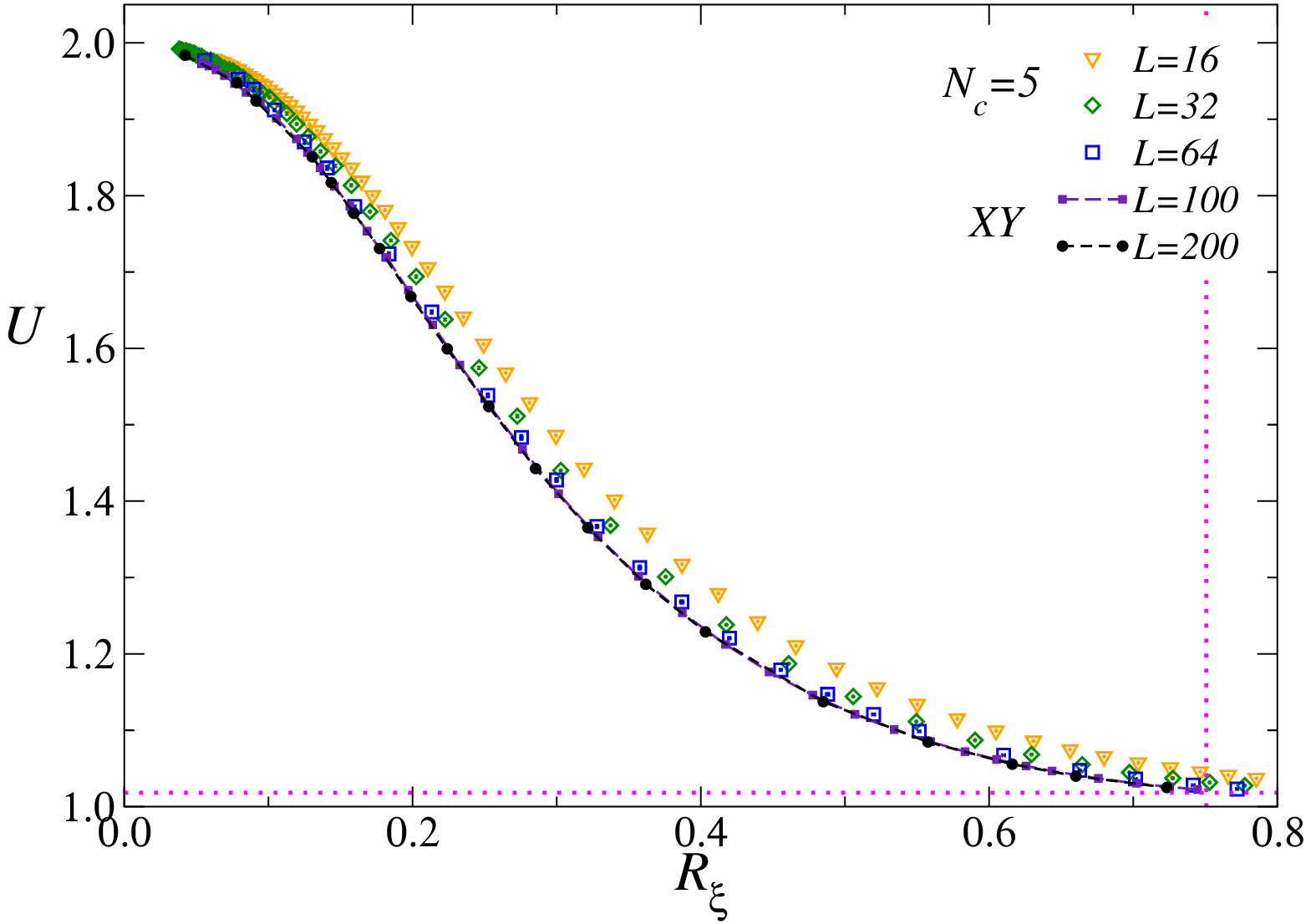}
\includegraphics[width=0.95\columnwidth, clip]{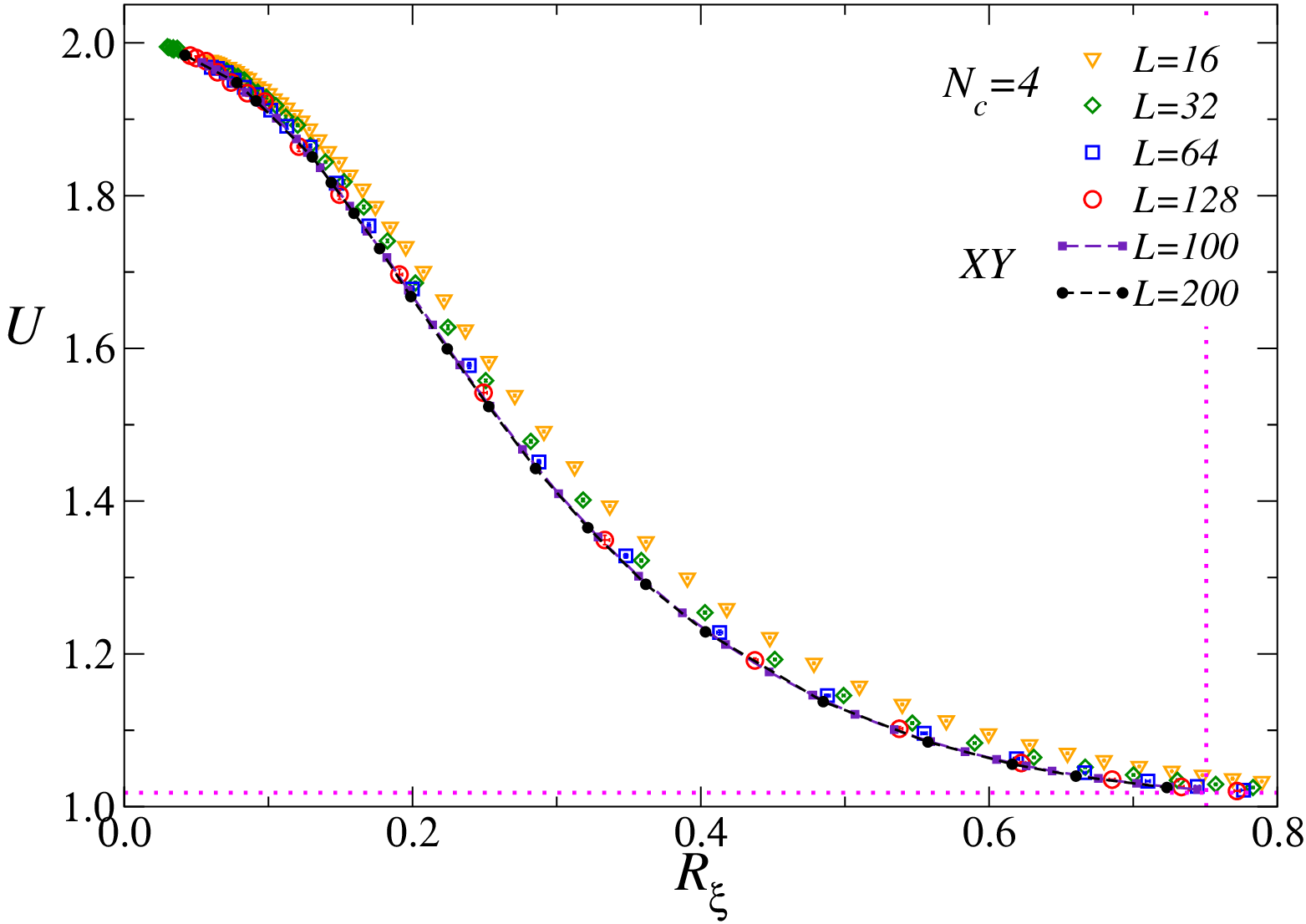}
  \caption{Plot of $U$ versus $R_\xi$ for $N_c=4$ (bottom) and $N_c=5$
    (top), at $\gamma=0$. We also report data for the 2D $XY$ model
    (\ref{xymodel}).  The FSS curve of the $XY$ model is clearly
    approached by the data for the lattice SO($N_c$) models with
    increasing $L$.  The horizontal and vertical lines indicate the
    BKT values $U(T_c)=1.018192(6)$ and $R_\xi(T_c) = 0.750691...$,
    respectively~\cite{Hasenbusch-05,Hasenbusch-08}.}
  \label{urxi45}
\end{figure}

To begin with, in Fig.~\ref{xibetanc3} we show the estimates of the
correlation length $\xi$ for $N_c=3$ and $\gamma=0$. They show a
sudden increase around $\beta\gtrsim 3.5$, as expected in the presence
of a finite-temperature BKT transition. To characterize the nature of
the transition, we plot the Binder parameter $U$ versus the ratio
$R_\xi$, In the FSS limit data should belong to a curve that only
depends on the universality class. In Fig.~\ref{urxi3} we report our
numerical results for $N_c=3$ and for three values of $\gamma$, which
are $\gamma=0,\,\pm 1$. In all cases, data appear to approach a
universal FSS curve with corrections that decrease quite rapidly with
the size.  We also report data for the 2D $XY$ model, that have been
obtained by standard Monte Carlo simulations for lattice sizes
$L=100,\,200$. They are apparently sufficient to provide a good
approximation of the asymptotic FSS behavior (the differences between
the $L=100$ and $L=200$ scaling curves are very small and hardly
visible in Fig.~\ref{urxi3}).  It is quite clear that the data for the
gauge model fall on top of the $XY$ scaling curve, confirming that the
transition has the same universal features: the gauge SO(3) model
undergoes a BKT transition as the $XY$ model.  Analogous results are
obtained for $N_c=4$ and $N_c=5$, as shown in Fig.~\ref{urxi45}, where
we report data for $\gamma=0$.  In both cases, the data for the gauge
model converge toward the FSS curve of the $XY$ model.

We note that the approach to the asymptotic FSS behavior
(\ref{r12sca}) is apparently quite fast in all lattice models
considered, including the 2D $XY$ model.  In particular, the scaling
corrections for the lattice SO($N_c$) gauge models appear to
effectively decrease roughly as $L^{-1}$ in the limited range of $L$
that we consider, up to $L=128$. At BKT transitions, logarithmic
corrections are generally expected
~\cite{HMP-94,HP-97,Hasenbusch-08,PV-13,CNPV-13}. However, our range
of values of $L$ is too small to allow us to detect logarithmic
changes of the estimates. In the range we consider power-law
corrections effectively dominate. Significantly larger sizes are
needed to allow us to perform fits that include both logarithmic and
power-law corrections.  Even though our analyses are not sensitive to
the slowly-decaying logarithmic corrections, we can argue that the
systematic error they induce is small (we only refer here to the
behavior of $U$ versus $R_\xi$; we are not claiming that logarithmic
corrections are always negligible). Indeed, the coefficients of the
logarithmic corrections are not universal, and therefore we expect
different logarithmic corrections in the $XY$ model and in the gauge
models we consider here. Thus, assuming that all models have a common
universal asymptotic behavior, we can infer the size of the
logarithmic correction by looking at the differences between the
results obtained in the different models. As apparent from
Figs.~\ref{urxi3} and~\ref{urxi45}, differences are tiny, indicating
that these elusive corrections play little role here.

Accurate estimates of the critical BKT temperatures are hard to
obtain, since their determination is generally affected by logarithmic
corrections, see Eq.~(\ref{rasym}).  The problem of the logarithmic
corrections can be overcome by the so-called matching method put
forward in Refs.~\cite{HMP-94,HP-97,Hasenbusch-05} (see also
Refs.~\cite{CNPV-13,DV-17} for applications to some 2D quantum lattice
gas models).  Here, we do not pursue this analysis further, since we
are not particularly interested in obtaining precise estimates of the
critical temperatures. We only mention some rough estimates of the
transition temperatures obtained by looking at the $\beta$-values
where $R_\xi(\beta,L)\approx R_\xi(T_c)=0.750691...$.  For $N_c=3$ we
find $\beta_c\approx 3.82$ for $\gamma=0$, $\beta_c\approx 3.77$ for
$\gamma=1$, and $\beta_c\approx 3.92$ for $\gamma=-1$.  Moreover, we
estimate $\beta_c\approx 4.80$ for $N_c=4$ and $\beta_c\approx 5.76$
for $N_c=5$, at $\gamma=0$.

In conclusion, the FSS analysis has allowed us to determine the nature
of the finite-temperature transitions occurring in the lattice
SO($N_c$) gauge model (\ref{hgauge}) with two flavors. For
$N_c=3,\,4,\,5$ we find that the transition belongs to the BKT
universality class, as in the classical $XY$ model. This occurs at
least in an interval of values of $\gamma$ around the infinite
gauge-coupling value $\gamma=0$.

\section{Conclusions}
\label{conclu}

We have studied a class of 2D lattice non-Abelian SO($N_c$) gauge
models with two real scalar fields, defined by the Hamiltonian
(\ref{hgauge}). Such lattice gauge models are obtained by partially
gauging a maximally O($2N_c$)-symmetric multicomponent real scalar
model, using the Wilson lattice approach. For $N_c \ge 3$, the
resulting theory is locally invariant under SO($N_c$) gauge
transformations and globally invariant under Abelian O(2)
transformations.  This study extends previous work on 2D models with a
local gauge invariance and a global non-Abelian
symmetry,~\cite{BPV-19-ah2,BPV-20-qcd2,BFPV-20-ong}, in which a
critical behavior can only be observed in the zero-temperature limit.
In the models considered here, instead, the global Abelian O(2)
symmetry may allow finite-temperature BKT transitions between the
disordered phase and the low-temperature QLRO phase.

The universal features of the transitions have been determined by
performing FSS analyses of Monte Carlo data.  We present results for
the two-flavor lattice SO($N_c$) gauge models (\ref{hgauge}) with
$N_c=3,4,5$.  They show that these systems undergo a
finite-temperature BKT transition that separates the disordered phase
from the low-temperature phase. Moreover, we have verified that the
low-temperature phase is characterized by spin waves, analogously
to the standard $XY$ model.

These results provide additional evidence in favor of the conjecture
that the critical behavior of 2D lattice gauge models, defined using
the Wilson approach~\cite{Wilson-74}, belongs to the universality
class of the field theories associated with the symmetric spaces that
have the same global symmetry. This conjecture assumes that gauge
correlations are not critical and decouple in the critical limit.
Therefore, the conjecture may fail when the gauge correlations are
critical, giving rise to a more complex behavior.  A similar
phenomenon has been observed in the three-dimensional lattice
Abelian-Higgs model with noncompact gauge fields, see, e.g.,
Ref.~\cite{BPV-20-ncah} and references therein.

We finally mention that the interplay between global and gauge
symmetries has also been studied in three dimensional models, see
Refs.~\cite{BPV-19-qcd3,BPV-20-on,PV-19,BPV-20-ncah}.

\bigskip

\emph{Acknowledgement}.  Numerical simulations have been performed on
the CSN4 cluster of the Scientific Computing Center at INFN-PISA.


\begin{thebibliography}{99}

\bibitem{Wilson-74} K. G. Wilson, Confinement of quarks, Phys. Rev. D
  {\bf 10}, 2445 (1974).

\bibitem{ZJ-book} J. Zinn-Justin, 
  {\em Quantum Field Theory and Critical Phenomena}, 
  fourth edition (Clarendon Press, Oxford, 2002).

\bibitem{Weinberg-book} S. Weinberg, {\em The Quantum Theory of
  Fields}, (Cambridge University Press, 2005).

\bibitem{Sachdev-19} S. Sachdev, Topological order, emergent gauge
  fields, and Fermi surface reconstruction, Rep. Prog. Phys. {\bf 82},
  014001 (2019).

\bibitem{Anderson-book} P.~W.~Anderson, {\em Basic Notions of
  Condensed Matter Physics}, (The Benjamin/Cummings Publishing
  Company, Menlo Park, California, 1984).

\bibitem{BPV-19-ah2} C. Bonati, A. Pelissetto and E. Vicari,
  Two-dimensional multicomponent Abelian-Higgs lattice models,
  Phys. Rev. D {\bf 101}, 034511 (2020).

\bibitem{BPV-20-qcd2} C. Bonati, A. Pelissetto, and E. Vicari,
  Universal low-temperature behavior of two-dimensional lattice scalar
  chromodynamics, Phys. Rev. D {\bf 101}, 054503 (2020).

\bibitem{BFPV-20-ong} C. Bonati, A. Franchi, A. Pelissetto, and
  E. Vicari, Asymptotic low-temperature critical behavior of
  two-dimensional mulitiflavor lattice SO($N_c$) gauge theories,
  Phys. Rev. D {\bf 102}, 034512 (2020).


  
\bibitem{MW-66} N.~D.~Mermin and H.~Wagner, Absence of ferromagnetism
  or antiferromagnetism in one- or two-dimensional isotropic
  Heisenberg models, Phys. Rev. Lett.  {\bf 17}, 1133 (1966).

\bibitem{BHZ-80} E. Br\'ezin, S. Hikami, and J. Zinn-Justin,
  Generalized non-linear $\sigma$-models with gauge invariance,
  Nucl. Phys. B {\bf 165}, 528 (1980).

\bibitem{BFPV-20} C. Bonati, A. Franchi, A. Pelissetto, and
  E. Vicari, Asymptotic low-temperature behavior of two-dimensional
  RP$^{N-1}$ models, Phys. Rev. D {\bf 102}, 034513 (2020).

\bibitem{KT-73} 
J. M. Kosterlitz and  D. J. Thouless,
  Ordering, metastability and phase transitions in two-dimensional systems,
  J.\ Phys. C: Solid State {\bf 6},  1181 (1973).

\bibitem{Berezinskii-70} V. L. Berezinskii, Destruction of Long-range
  Order in One-dimensional and Two-dimensional Systems having a
  Continuous Symmetry Group I. Classical Systems,
  Zh. Eksp. Theor. Fiz. {\bf 59}, 907 (1970) [Sov. Phys. JETP {\bf
      32}, 493 (1971)].

\bibitem{Kosterlitz-74}
  J. M. Kosterlitz, 
  The critical properties of the two- dimensional xy model,
  J. Phys. C {\bf 7}, 1046 (1974).

\bibitem{JKKN-77} J. V. Jos\'e, L. P. Kadanoff, S. Kirkpatrick, and
  D. R. Nelson, Renormalization, vortices, and symmetry-breaking
  perturbations in the two-dimensional planar model, Phys. Rev. B
  {\bf 16}, 1217 (1977).

\bibitem{HMP-94} M. Hasenbusch, M. Marcu, and K. Pinn, High precision
  renormalization group study of the roughening transition, Physica A
  {\bf 208}, 124 (1994).

\bibitem{HP-97} M. Hasenbusch and K. Pinn, Computing the roughening
  transition of Ising and solid-on-solid models by BCSOS model
  matching, J. Phys. A {\bf 30}, 63 (1997).

\bibitem{Balog-01} J. Balog, Kosterlitz-Thouless theory and lattice
  artifacts, J. Phys. A {\bf 34}, 5237 (2001).

\bibitem{Hasenbusch-05} M. Hasenbusch, The two dimensional XY model at
  the transition temperature: a high precision numerical study,
  J. Phys. A {\bf 38}, 5869 (2005).

\bibitem{PV-13} A. Pelissetto and E. Vicari, Renormalization-group
  flow and asymptotic behaviors at the Berezinskii-Kosterlitz-Thouless
  transitions, Phys. Rev. E {\bf 87}, 032105 (2013).

\bibitem{BPV-20-on} C. Bonati, A. Pelissetto, and E. Vicari,
  Three-dimensional phase transitions in multiflavor scalar SO($N_c$)
  gauge theories, Phys. Rev. E {\bf 101}, 062105 (2020).

\bibitem{PV-02} A. Pelissetto and E. Vicari, Critical phenomena and
  renormalization group theory, Phys. Rep. {\bf 368}, 549 (2002).

\bibitem{FB-72} M. E. Fisher and M. N. Barber, Scaling theory for
  finite-size effects in the critical region, Phys. Rev. Lett. {\bf
    28}, 1516 (1972).

\bibitem{Barber-83}
  M. N. Barber, in {\em Phase Transitions and Critical Phenomena},
  edited by C. Domb and J. L. Lebowitz (Academic Press, New York, 1983),
  Vol. 8.

\bibitem{Privman-90} V. Privman ed., {\em Finite Size Scaling and
  Numerical Simulation of Statistical Systems} \/ (World Scientific,
  Singapore, 1990).

\bibitem{Hasenbusch-08}
M. Hasenbusch,  The Binder cumulant at the Kosterlitz-Thouless transition,
J. Stat. Mech.: Theory Expt. P08003 (2008).


\bibitem{CNPV-13} G. Ceccarelli, J. Nespolo, A. Pelissetto, and
  E. Vicari, Universal behavior of two-dimensional bosonic gases at
  Berezinskii-Kosterlitz-Thouless transitions, Phys. Rev.  B {\bf 88},
  024517 (2013).

\bibitem{DV-17} F. Delfino and E. Vicari, Dimensional crossover of
  Bose-Einstein condensation phenomena in quantum gases confined
  within slab geometries, Phys. Rev. A {\bf 96}, 043623 (2017).

  \bibitem{Cabibbo:1982zn} 
  N.~Cabibbo and E.~Marinari,
  A New Method for Updating SU(N) Matrices in Computer Simulations of Gauge Theories,
  Phys.\ Lett.\  {\bf 119B}, 387 (1982).

\bibitem{Metropolis:1953am} N.~Metropolis, A.~W.~Rosenbluth,
  M.~N.~Rosenbluth, A.~H.~Teller, and E.~Teller, Equation of state
  calculations by fast computing machines, J.\ Chem.\ Phys.\ {\bf 21},
  1087 (1953).

\bibitem{Creutz:1987xi} M.~Creutz, Overrelaxation and Monte Carlo
  Simulation, Phys.\ Rev.\ D {\bf 36}, 515 (1987).

\bibitem{CFT-book}
P. Di Francesco, P. Mathieu, and D. Senechal,
{\em Conformal Field Theory} (Springer Verlag, New York, 1997).

\bibitem{HPV-05} M. Hasenbusch, A. Pelissetto, and E. Vicari,
  Multicritical behaviour in the fully frustrated XY model and related
  systems J. Stat. Mech.: Theory Expt.  P12002 (2005).

\bibitem{BPV-20-ncah} C. Bonati, A. Pelissetto, and E. Vicari, Lattice
  Abelian-Higgs models with noncompact gauge field, arXiv:2010.06311.

\bibitem{BPV-19-qcd3} C. Bonati, A. Pelissetto and E. Vicari, Phase
  diagram, symmetry breaking, and critical behavior of
  three-dimensional lattice multiflavor scalar chromodynamics,
  Phys. Rev. Lett. {\bf 123}, 232002 (2019); Three-dimensional lattice
  multiflavor scalar chromodynamics: interplay between global and
  gauge symmetries, Phys. Rev. D {\bf 101}, 034505 (2020).

\bibitem{PV-19} A.~Pelissetto and E.~Vicari, Multicomponent compact
  Abelian-Higgs lattice models, Phys. Rev. E {\bf 100}, 042134 (2019).


\end{thebibliography}
\end{document}